\documentstyle[psfig]{l-aa}

\def\etal{{\it et al.~}}

\def\vhbm{$V_{\rm HB}$}
\def\vhb{$V_{\rm HB}$}
\def\vzahb{$V_{\rm ZAHB}$}
\def\C12C13{$^{12}$C/$^{13}$C}
\def\bmvg{$(B-V)_{o,g}$}
\def\d14{$\Delta V_{1.4}$ }
\def\fe{[Fe/H]~}
\def\hst{{\it HST~}}

\begin{document}

\thesaurus{
(08.16.3, 08.08.1, 10.07.3, )} 
\title{ THE STELLAR POPULATION OF THE GLOBULAR CLUSTER M 3}

\subtitle{ II. CCD PHOTOMETRY OF ADDITIONAL 10,000 STARS  }

\author{ F.R. Ferraro\inst{1,2}, E. Carretta\inst{1,2}, C.E. Corsi\inst{3}, 
F. Fusi Pecci\inst{1}, C. Cacciari\inst{1}, R. Buonanno\inst{3},
B. Paltrinieri\inst{4}, D. Hamilton\inst{5}}

\institute{\inst{1} Osservatorio Astronomico di Bologna, Via Zamboni 33, 
I-40126 Bologna, ITALY\\
\inst{2} Visiting Astronomer, German-Spanish Astronomical Center, Calar Alto, 
SPAIN\\
\inst{3} Osservatorio Astronomico di Roma, Roma, ITALY\\
\inst{4} Dipartimento d'Astronomia, Via Zamboni 33, I40126 Bologna, ITALY\\
\inst{5} Max-Planck Institut f. Astrophysik, Heidelberg, GERMANY}

\offprints{ F.R. Ferraro }

\date{}

\maketitle
\markboth{F.R.~Ferraro et al.}{}

\begin{abstract}

We present BVI CCD photometry for more than 10,000 stars in the innermost
region (0.3' $<$ r $<\sim$ 4') of the globular cluster M 3. 
When added to the previous photographic photometry by Buonanno \etal (1994) 
reaching as far as r $\sim$ 7', this results in an homogeneous data-set
including about 19,000 stars measured in this cluster, which can be now
regarded as one of the main templates for stellar population studies.

Our main results, from the new colour-magnitude diagrams (CMD), are: 
\par\noindent
\begin{enumerate}

\item 
Completeness has been achieved for all objects at V$\le$18.6 
and $0.3' < r < 7'$. 

\item 
Our new independent photometric calibration is redder than the old
calibration (Sandage and Katem 1982) in the blue range, and bluer 
in the reddest part of the CMD. This colour term has important consequences 
on some issues (e.g. metallicity, period-shift effect, etc.). 

\item 
The metallicity derived from the photometric indicators $(B-V)_{0,g}$ 
and $\Delta V_{1.4}$ is [Fe/H]$\sim$-1.45, i.e. significantly higher than the 
value -1.66 generally used so far. This is in very good agreement with the 
most recent high resolution spectroscopic studies of individual stars. 

\item
There is indication of the presence of a very faint and blue extension 
of the HB, although scarcely populated. 

\item
The population ratios of HB, RGB and AGB stars confirm the value previously 
found for the helium content, i.e. Y=0.23$\pm$0.02. 

\item
There is a significant population of blue straggler stars, although the 
exact number must await HST data for a better resolution in the very
central regions.

\end{enumerate}

\end{abstract}

%

\section{Introduction}

The advent of the charge-coupled devices technology and, very recently,
the availability of the refurbished {\it Hubble Space Telescope} (\hst), 
coupled with the use of more and more sophisticated software for photometric 
analysis in crowded fields, have opened unique possibilities of using 
galactic globular clusters (GGCs) as templates for testing the stellar 
evolution theories (Renzini and Fusi Pecci 1988--RFP88). 
In particular, the combined use of both colour-magnitude diagrams (CMDs) 
and luminosity functions (LFs) derived from very accurate and complete
photometric studies allows to reach levels of precision eventually 
suitable to check even the finer details predicted from updated theoretical 
models.

Since the basic aim is to secure stellar samples {\it as populous and complete
as possible in any radial region} of a cluster to yield statistically
significant information on all the stellar evolutionary phases (including the 
very short ones, $t\sim 10^4-10^5$ yr) and on the cluster as a whole 
(f.i. taking 
into account also the cluster internal dynamics), proper ground-based 
observations in the outer areas must complement \hst observations exploiting 
the photometry in the very central regions.

Within this framework, we started a long-term project (ground-based + \hst)
to make M3 {\it one of the first and best observed templates} to verify 
in detail the model predictions (Buonanno \etal 1986, 1994, RFP88, 
Ferraro \etal 1993, Cacciari \etal 1993).

Why should M3 justify such a big effort? 
There are many good reasons. Among others, for instance: a) since
the early study by Sandage (1953), M3 is one of the proto-type Pop II
clusters of intermediate-poor metallicity; b) it is rich of RR Lyrae
variables, and is usually adopted as Oosterhoff (1939) class I
prototype; c) it contains the first detected {\it blue straggler stars}
(BSS), now suspected to have a bimodal radial distribution (Ferraro
\etal 1993); d) its CMD displays a significant population in every
branch, with an Horizontal Branch (HB) spanning a very wide range
in color (temperature), including extremely blue (hot) stars and
Post-Asymptotic Giant Branch objects (P-AGB); e) it has been the
target of recent spectroscopic studies (including at high resolution,
see for references Carretta and Gratton 1996a), yielding a new
insight on its average metallicity and metallicity dispersion.

The {\it first} step in our project has been a wide-field survey, based on the
original Sandage's photographic plates, of regions from 2 up to 7 arcminutes 
from the cluster center, whose results have already been published 
(Buonanno \etal 1994--PH94). In that paper we published $B$,$V$ magnitudes
for more than 
10,000 stars, reaching about two magnitudes fainter than the 
Main-Sequence turnoff (TO), with internal photometric errors mostly smaller 
than 0.05 mag. As a {\it second step}, in this paper  we present 
the results of a new BVI CCD-photometry for almost 10,000
additional stars measured in the inner parts of the cluster. The {\it third}
and final step to complete the survey will present the \hst data secured
on the cluster core (proposal GO 5496, P.I. F. Fusi Pecci), whose reduction is 
currently in progress (Fusi Pecci \etal 1996). A further paper specifically
devoted to the BVI CCD-photometry of 65 RR Lyrae variables is in preparation
(Cacciari \etal 1996).

In Section 2, we present the description of data acquisition and analysis, 
including tests for completeness and comparisons with our previous 
photographic results (PH94) and with other 
data-sets. In Section 3, we describe our results, with special emphasis 
on the features of the main bright branches of the CMD and, in particular,
we deal with mean ridge lines, the so-called RGB-Bump, star counts, 
population ratios, and an update of the overall blue stragglers population 
in M 3. 
Finally, Section 4 is devoted to conclusions and schematic summary.

\section{Observations, data reduction, comparisons and completeness}

\subsection{Observations and cluster sampling}

The CCD data were obtained at the 3.6m CFHT telescope on April 1991
by H. Richer and R. Buonanno within a wider collaborative program, using a
2,048 $\times$ 2,048 pixels detector with field of view of $\sim$ 7' $\times$
7'. 
The observations consisted of two sets of BVI frames reported in
Table~\ref{t:obslog}. The short exposures were specifically taken to survey the
very inner regions, avoiding strong saturation effects.
Since the detector had a very poor response in the blue, 
Ferraro \etal (1993--F93) used the deepest $V$ and $I$ exposures
to get a first hint on the BSS population in the cluster 
central regions ($r <$2').

\begin{table}
\caption{ Observation log of the BVI CCD data}
\begin{tabular}{lccc}
\hline\hline
\\
Colour   & Exp.Time &Date & Seeing \\
	 &   (sec)  &     & (arcsec)\\   
\hline
\\
B        &  120    & April 7,1991 & 0.80 \\
V        &  120    & April 7,1991 & 0.65 \\
I        &  120    & April 7,1991 & 0.63 \\
B        &  30     & April 7,1991 & 0.85 \\
V        &  10     & April 7,1991 & 0.70 \\
I        &  10     & April 7,1991 & 0.63 \\
\\
\hline
\end{tabular}
\\
\label{t:obslog}
\end{table}

Figure~\ref{f:f1} shows the position of the area surveyed in the present study 
(the {\it full square}) overlapped to the map of the regions considered in 
the photographic study (PH94). For sake of clarity, the individual
stars measured on the plates in the PH94 bright ({\it crosses})  and 
faint ({\it small dots}) sample are respectively plotted.

\begin{figure}[htbp]
\caption{ Map of the regions of M 3 covered with the photographic
survey (PH94). The {\it full square} reproduces the field 
covered by the present CCD-observations. {\it Crosses}
indicate stars included in the PH94 {\it bright sample},
while {\it dots} indicate stars membering the photographic {\it
faint sample}. In the map, North is up and east is to the right;
the scale-unit is in arcsecond.} 
\label{f:f1}
\end{figure}

To give a global overview of the planned survey,
we report in Figure~\ref{f:f2} also the size and location of the three 
WFPC2 fields we observed with GO 5496. As it can be seen, there is
a useful overlapping between the photographic, CCD, and \hst considered
areas. For comparison, we have also reported in the same figure ({\it dashed 
squares}) the fields covered by other recent photometric studies on M 3, 
namely those by Bolte \etal (1995; hereinafter BHS), Guhathakurta \etal
(1994; hereinafter GYBS), and Burgarella \etal (1995)
using the WFPC of \hst before refurbishment.

\begin{figure}[htbp]
\caption{ Global overview of the fields observed in the photographic,
CCD, and \hst surveys that will compose our overall photometric data-set 
in M 3. The {\it full square} represents the contour of the CFHT field.
The {\it dashed squares} are the fields surveyed by Bolte \etal
(1995; label B.) and  Guhathakurta \etal (1995; label G.), respectively.
Note that the FOC-field observed by Burgarella \etal (1995) would be
represented here by just a ``central dot''.
The annulus (2.0' $< r <$ 3.5'; {\it thin solid lines}) is the 
region of M 3 where the ``complete'' sample considered in the photographic
study has been obtained (PH94). The small {\it full circle} represents
the central region (with $r<20"$) where crowding makes unfeasible any
meaningful photometry on the deep CCD-frames taken at the  CFHT.}
\label{f:f2}
\end{figure}

\subsection{Data reductions}

The data analysis was performed in Bologna and Rome using ROMAFOT (Buonanno \etal
1979, 1983), whose standard procedures have been reported in
other papers (see for references Ferraro \etal 1990, F93). No special
requirement has been involved either in the searching phase or in
measuring the instrumental magnitudes. 

Since the only wide set of standard stars available in the M3 field 
we observed were those
measured with the photoelectric photometer by Sandage (1953, 1970), 
later revised by Sandage and Katem (1982), we put some effort to obtain an 
independent CCD-calibration. As a matter of fact, neither BHS (they 
only presented instrumental magnitudes) nor GYBS (who ultimately referred
their magnitudes to Auriere and Cordoni 1983 
and F93) obtained an independent calibration.
The photometry of Auriere and Cordoni (1983) in the central regions
was linked to star profiles (King 1966) and cannot be used 
for calibration. The data presented by Scholz and
Kharchenko (1994) are referred to Sandage (1970). Finally, Paez \etal 
(1990) in their CCD-study of the outer regions (at about 5 arcmin
from the cluster center) did not get any independent calibration but
referred their data to 41 stars in common with Sandage and Katem (1982).

In PH94,  we used the sequence of photoelectric standards from Sandage 
(1970) to calibrate the photographic magnitudes to the standard $B,V$ Johnson 
system. In F93, the CFHT data were calibrated using two standard fields 
observed during the same nights in NGC 4147 and M 67, with the well known 
problems related to the difficult handling of calibrated stars in
crowded standard fields. Eventually, we concluded that a new independent
CCD-calibration was necessary starting from a proper set of primary
standards.

After several unfruitful runs (due to bad weather, telescope failures etc.),
our new calibration is based on an observing run carried out at the 1.23 m 
telescope at the German-Spanish Astronomical Centre, Calar Alto, 
Spain. BVI CCD-frames of M 3 and of several Landolt CCD standard areas were
acquired under photometric conditions in the night of April 2, 1995, using a
thinned 1024 $\times$ 1024 Tektronix chip (24 $\mu$ pixels, 0.50 arcsec/pixel), 
with Ar-coatings. The $I$ filter is in the Kron-Cousin system, centered at
8020~\AA.

The calibration of the CFHT CCD magnitudes was thus performed in two steps.
First, we used 18 stars in 7 CCD standard areas (Landolt 1983) to 
link the Calar Alto instrumental magnitudes to
the standard Johnson system. The final equations we adopted for 
the B,V,and I filters are reported below and plotted in Figure~\ref{f:f3}: 
\begin{equation}
\label{eq:1}
{\rm B} = {\rm b} + 0.215\,(\pm 0.017) {\rm (b - v)} + 20.386\,(\pm 0.007) 
\end{equation}
($\sigma=$0.029),

\begin{equation}
\label{eq:2}
{\rm V} = {\rm v} + 20.577\,(\pm 0.005)
\end{equation}
($\sigma=$0.020),

\begin{equation}
\label{eq:3}
{\rm I} = {\rm i} + 20.340\,(\pm 0.006)
\end{equation}
($\sigma=$0.026),

\noindent
where b,v,i, are the instrumental magnitudes.\footnote{Throughout this 
paper, the symbol $\sigma$ will indicate the standard deviation of a 
single measurement, while the value after the symbol $\pm$ will refer 
to the standard deviation of the mean.} 
The atmospheric extinction terms in each colour have been derived by
observing the same standard field at different airmass during the night.

\begin{figure}[htbp]
\caption{ Calibrating Landolt CCD standard stars used to transform the 
instrumental Calar Alto magnitudes to the Johnson standard system.
Different symbols refer to stars in different standard areas.}
\label{f:f3}
\end{figure}

In the $I$ calibration we have used only the measurements having a 
sufficient S/N. In particular, this has led to exclude two of the bluest
standard. because of this, we could exclude the existence of residual colour
terms for the $I$ band.

These equations have been used to calibrate the instrumental magnitude of a
sample of $\sim$ 100 isolated, unsaturated stars in the Calar Alto frames 
for which the aperture photometry has been performed. This set of stars has 
been used to correct the CFHT magnitudes (previously calibrated with the
NGC 4147 and M 67 standard fields, see F93). The relations linking the ``old"
CFHT calibration system to the new standard system (CCD96) are:

\begin{equation}
\label{eq:4}
{\rm V_{CCD96}} = {\rm V_{CFHT}} + 0.02 {\rm (B - V)_{CFHT}} - 0.004
\end{equation}

\begin{equation}
\label{eq:5}
{\rm B_{CCD96}} = {\rm B_{CFHT}} + 0.03 {\rm (B - V)_{CFHT}} - 0.006
\end{equation}

\begin{equation}
\label{eq:6}
{\rm I_{CCD96}} = {\rm I_{CFHT}} - 0.03 {\rm (V - I)_{CFHT}} + 0.054
\end{equation}

Following the above procedure we linked the whole CFHT-CCD data set to the 
Johnson system. In the next section we find the relation to transform the 
photographic data set to the new standard system here adopted.

In this respect, it is important to note that, after completion of our
reductions, we became aware of the existence of a new independent
photometric study  of a very wide sample of M3 stars (about 23,700) carried
out by K.A. Montgomery (1995 -- M95) as part of his Ph.D. dissertation,
based on 2048$\times$2048 (T2ka) CCD-frames taken with the 0.9-meter 
telescope at KPNO.
Since with the f/7.5 secondary, the telescope produced an image scale of
0.68 arcsecs/px and a field of view of 23.2 arcminutes, and moreover
2$\times$2 mosaics of images were made with the centers offset by 10
arcminutes in both right ascension and declination, the total field
covered by Montgomery fully overlaps both our field and the outer
field independently observed by Stetson and Harris (1988), which being
located outside the region we observed could not be used in our
photometry. We will discuss in detail the results of the comparisons
with this new data-set at Sect. 2.4 and 2.5.

\subsection{ Photographic $vs.$ CCD photometry; 
magnitudes and positions for the global sample of M 3}

More than 900 stars have been found in common between the CCD96 and PH94
(bright and faint) sample. However, due to the limited performances of the used
CCD detector in the blue band, the photometric internal errors in the $B$
band are large and reduce the reliability
of the transformations. For this reason only 532 bright ($B<18.6$)
stars in common have been used to derive the relations linking the photographic
photometric system to the standard system here adopted.
The final relations correcting the $V$ magnitude and the $B-V$ colours are: 

\begin{equation}
\label{eq:7}
{\rm V_{CCD96}} = {\rm V_{PH94}} + 0.08 {\rm (B - V)_{PH94}} + 0.053
\end{equation}

\begin{equation}
\label{eq:8}
{\rm (B-V)_{CCD96}} = {\rm (B-V)_{PH94}} -0.143 {\rm (B - V)_{PH94}} + 0.096
\end{equation}
\noindent

These corrections have been applied to the whole (bright + faint) PH94 sample. A
small residual non-linearity in the photographic data (probably due to the
original plate transformation) has been corrected using a polynomial fit of
5$^{th}$ order (to yield $C^{5th}_V$ and $C^{5th}_B$ for the $V$ and $B$ filter,
respectively). This additional correction is properly determined 
only for bright stars ($B<16.8$, $V \sim 17.8$). 

Resting on our own data alone, it would be impossible to check properly the
linearity of the fainter PH94 measures. However, by comparing our data with the
quite similar data-set obtained by M95, we can extend the linearity test to
fainter magnitudes, as explained in Sect. 2.4. From this last
comparison we have obtained further (small) corrections for the fainter stars
(see Sect. 2.4) using a similar procedure as adopted for the bright sample. 

Then, to avoid any discontinuity at the junction between the bright 
and the faint PH94 samples, we have imposed coincidence between 
the corrections computed at the junction [$C^{5th}_{V=16.8}$ and 
$C^{5th}_{B=17.8}$] from the two different CCD data-sets.
Fortunately, they actually turned out to be almost equal, and this 
reinforces both the validity of the procedure and the achievement of
a good linearity allover the ``revised'' PH94 sample.

In Figure~\ref{f:f4} and Figure~\ref{f:f5},  we present
the ($V,B-V$) and ($V,V-I$) CMDs containing only stars outside a radius of 
20" from the cluster centre, excluding the variables which are however
included in the lists. Due to
the low efficiency in the blue of the CFHT CCD-camera we have plotted
in Figure~\ref{f:f4} only the objects measured on our B CCD images
with $B<18.6$. 
In particular, the upper part of the 
$V,B-V$ CMD (brighter than  $B=18.6$) contains 2227 constant and 154 variable 
stars taken from the CFHT CCD sample, plus 42 variable stars and 444 constant 
stars 
taken from the {\it bright} photographic sample outside the CCD field.
For magnitudes fainter than $B=18.6$, we report the 9339 stars
from the {\it faint} PH94 sample. The total number of stars in 
this CMD is therefore 12206.
The $V, V-I$ colour magnitude diagram contains 9647 stars 
including 155 variables that have both $V$ and $I$ magnitudes. 

\begin{figure}[htbp]
\caption{ CCD96 $V, B-V$ colour-magnitude diagram for 12206
stars in M 3 with r $>$ 
20". Variable stars are not shown; only stars with $B<18.6$ from the CFHT 
sample have been plotted. See text for details.}
\label{f:f4}
\end{figure}

\begin{figure}[htbp]
\caption{ CCD96 $V, V-I$ colour-magnitude diagram for 9647
stars in M 3 with r $>$ 
20". Variable stars are not included in the plot, see text for details.}
\label{f:f5}
\end{figure}

All the available data (magnitudes and positions) are available upon request
from the authors, and will also be sent to Strasbourg Data Center. As in the
PH94 paper, X, Y coordinates (in arcsec) are referred to the cluster centre,
taken at $\alpha_{1950} = 13^h 39^m 24^s$ and $\delta_{1950} = + 28^0 38'$.
Coordinate transformation to other systems can be easily performed; to help
with cross-identification, we note that our bright stars No. 8877, 10052 and
12045 are the stars No. 9, 3842 and 2267, respectively, in the photometric 
catalogue of GYBS. This allows conversion of our positions to equinox 2000 
coordinates, since GYBS's positions are in R.A. and Dec. referred
to the equinox 2000.

\subsection{Comparison with other studies}

We compare here our CCD-calibrated photometry with the data taken
from other studies for the stars in common. 

\smallskip\noindent
$\bullet$ {\it Comparison with Sandage and Katem 1982}
\par\noindent
The comparison was made using 81 stars from Tables I and II of SK82, cross-
identified with our data. The results of this comparison, together with the 
least squares fits through the data, are shown in 
Figure~\ref{f:f6}.

As one can see, the overall agreement is good, but there are also important
differences. First, there is a luminosity shift in both B and V,
variable with varying the B-V colour. Second, there is a quite noticeable
colour term, in the sense that  our data are $\sim$ 0.08 mag redder at 
$B-V=-0.2$ (the bluest point of the HB), $\sim$ 0.03 mag redder at 
$B-V=+0.3$ (corresponding to the middle of the instability strip), and 
$\sim$ 0.08 mag $bluer$ at $B-V=1.6$, $i.e.$ at the termination of the 
Red Giant Branch (RGB).

\begin{figure}[htbp]
\caption{ Differences between our calibrated magnitudes and those from Sandage
and Katem (1982) for 81 stars (extracted from his Tables I and II). In each
panel the equations of the least-squares fits (solid lines) to the data
are also reported.}
\label{f:f6}
\end{figure}

\smallskip\noindent
$\bullet$ {\it Comparison with Guhathakurta \etal 1995}
\par\noindent
Figure~\ref{f:f7} shows the results of a similar comparison with the $V$ and
$I$ magnitudes of GYBS for 634 stars, almost down to the completeness limit of
the survey. GYBS studied the central field of M 3 using the F555W
and F785LP bands of the {\it HST/Planetary Camera-1}, that are similar to the
Johnson $V$ and $I$ bandpasses, respectively. 

As explained below, this comparison is actually more useful for checking
completeness since, after a first colour transformation from $HST$ instrumental 
magnitudes, GYBS linked their photometry to the available ground-based
data. In particular, their final values include the application of zero-point 
offsets to the $V$ and $I$ data to match fiducial points of the upper parts 
of the CMDs presented by Auriere and Cordoni (1983) and 
$F93$. The generally good agreement shown Figure~\ref{f:f7} is thus
obvious though the F93 CCD calibration was still based on the NGC 4147 and
M67 standard areas.

\begin{figure}[htbp]
\caption{ Residuals between CCD96 $V$ and $I$ magnitudes and the corresponding
photometry of GYBS in the F555W ($V$) and F785LP ($I$) bands of 
$HST$/Planetary Camera-1 (see text). The comparison is based on 634 stars in
common within the area covered by GYBS.} 
\label{f:f7}
\end{figure}

\smallskip\noindent
$\bullet$ {\it Comparison with Montgomery 1995, and Stetson and Harris 1988}
\par\noindent 
Concerning the data-set obtained by M95, we present here 
the results of the comparisons made using the still preliminary list
of magnitudes and colors kindly made available to us by Dr. K. Janes.
Since they could still be subject to further analysis and revisions,
and will be eventually published in the near future (Janes
and Montgomery, 1996), we schematically report on two
main aspects which are of interest for the present study: the linearity test 
and the comparison of the zero-points.

Concerning the linearity test, after carrying out several different procedures,
we decided to restrict the sample to a sub-set of the 7,152
stars in common (from the RGB tip down to the faintest magnitude limit),
which have unambiguous identifications, no detected companions in both
photometries, and sufficiently low and well difined background. This implies
that these stars lie mostly in the outer cluster regions and have
good internal photometry in both studies. 

From the study of the residuals it is confirmed that, while no significant
trend is evident after the corrections applied to the bright PH94 stars
using our corresponding CCD data (see Sect. 2.3), there is still a small 
residual non-linearity in the faint photographic PH94 data, quite similar
in size and morphology to that detected for the bright stars with respect to 
our CCD measures. Therefore, we have corrected the faint PH94 measures
using again a polynomial fit of 5$^{th}$ order (computed over $\sim$ 3,800
faint stars) as done for the bright stars, and imposing coincidence of the 
(small, $\sim$0.01-0.02 mag) linearity corrections at the junction between 
the bright and faint samples (see Sect. 2.3).

The comparison of the zero-points has also been made on a special
sub-set of about 2,000 stars, chosen within the sample considered for
the linearity test but with V$<19.5$ (to avoid low S/N data).
The final figures for the differences, in the sense {\it M95 --
Ours}, are: $\Delta V=-0.06$, $\Delta B=-0.11$, $\Delta B-V=-0.05$
mag, respectively. As can be seen, especially for the B-magnitudes, the
residuals are quite large and somewhat worrying.

Based on the description reported in Table 2.1 and Fig. 2.3 of his thesis,
Montgomery observed Standard Star fields from Landolt (1992) over the six
night run at KPNO, securing approximately 250 measurements with internal
accuracy in the individual measure better than 0.05 mag. The KPNO run standard
deviations of the residuals he quoted were 0.018 in V and 0.024 mag in B,
quite comparable with the figures from our own calibration (see eqs.
(1)-(2)). It is therefore difficult to explain the differences between the
zero-points adopted in the two photometries based just on the formal errors.
And it is also difficult to choose which of the two may be correct. To have
further insight on this important aspect, we have carried out a further
comparison with the data obtained by Stetson and Harris (1988, --SH88).

Since there is no overlap between our surveyed regions and the area
observed by SH88, we must
use the comparisons (for different sets of stars) between M95 and
SH88 on the one hand, and M95 and this study, on the other, to provide
a link between SH88 and our photometry. So doing, we get that
the differences between our data and SH88 (in the sense {\it Ours --
SH88}) are: $\Delta V=+0.04$, $\Delta B=+0.04$, and 
$\Delta B-V=0.00$ mag. These figures are compatible
with the residuals measured by Montgomery with respect to SH88.
From Fig. 2.13 and Table 2.3 of Montgomery (1995)
one has in fact for the residuals {\it M95 -- SH88}
$\Delta V=-0.048$, $\Delta (B-V)=-0.020$ mag, based on 27 
stars in common with SH88. As a matter of fact, the zero-points of 
SH88 turn out to be intermediate between the other two photometric 
calibrations.

\subsection {Error estimates and choice of the zero-points }

Since we have only two frames in each colour with quite different 
exposure times, we cannot use repeated measures of the same star to 
yield a direct estimate of the {\it internal} measuring errors. 
However, we can get some quantitative hints
from the photometric parameters listed for each star within ROMAFOT (which 
essentially trace the quality of the fit) and from the comparison of the 
morphology of the individual branches in the CMD (at different magnitude
levels) with those obtained in previous studies. In particular,
using also as reference the error estimates discussed in PH94, we are 
fairly confident that the {\it internal} errors in the present photometry
are $\sim$0.02--0.03 mag at $V<18$, and $\sim$0.05 at the TO region.

Concerning the {\it systematic} errors, we have already discussed the
problems related to the zero-points and the colour transformations.
There are now three independent calibrations referred to observations
of Landolt's (1992) standards. They are in marginal agreement, but it
is difficult at the present stage to choose which is the best.

Since the calibration here adopted seems to be compatible with
that obtained by Stetson and Harris (1988) (they actually show
a systematic difference of 0.04 both in B and V, yielding however no colour
difference), we decided to keep using our calibration.

To be conservative and taking into account the size of the uncertainties, 
we conclude however that we still cannot exclude errors
as high as 0.05 mag in the {\it absolute} values, both in magnitude
and in colour (particularly at the blue and red extremes).

Such a conclusion is quite disappointing in particular for what
concerns the specific study of the RR Lyrae variables (Cacciari
\etal 1996), and further efforts to improve the absolute calibrations
are urged.

\subsection{Completeness of the CMDs}

Besides the comparisons in colour and magnitude discussed above, the BHS, GYBS,
and M95 samples can be also used to test in a quantitative way the degree of
completeness actually achieved in the present photometric data set. These
checks can be performed on different subsets, depending on the regions and the
magnitude range covered in the different studies. 

\smallskip\noindent
$\bullet$ {\it Comparison with Bolte \etal 1993}
\par\noindent
Of particular interest is the comparison with the BHS's samples, since they 
were obtained with the same CFHT telescope, `` under apparently photometric
conditions and with excellent seeing''. A careful check is feasible
using all stars brighter than $V=$ 16.75, the limit in $V$ 
used by BHS for the reference population in order 
to study  the blue stragglers distribution (see below).
The result of this check is plotted in Figure~\ref{f:f8} that compares 
the ($V$, $V-I$) CMD we have obtained from surveying the same area 
as BHS (see Figure~\ref{f:f2}) with their ($V$, $V-R$) diagram.
Note that their instrumental $V$ magnitudes have been transformed to our 
system by using the brightest (12.5 $< V <$ 17.0) 544 stars in common.

\begin{figure}[htbp]
\caption{ Comparison of the upper part of our CMD (panel b) with that of BHS 
(panel a) in the region covered by their study, for stars brighter than 
$V=$ 16.75. Instrumental $v$ magnitudes of BHS have been shifted to match 
our own calibrated $V$ system. Variable stars are here included and this
explains the ``fatness'' of the HB (see text).}
\label{f:f8}
\end{figure}

In the region in common between the two samples, BHS have found 801 stars
while we detected 831 objects, at $V<16.75$. Among these, 775 stars were
successfully cross-identified,  whereas 26 stars
from BHS and 53 stars from our survey had no mutual correspondence. However,
we must note that most of these residual stars lie on the faint RGB, just above
the threshold ($V=16.75$) adopted to carry out the comparison. Therefore, 
the difference in the samples essentially depends on small 
variations in the arbitrary offset necessary to match the two data-sets in 
$V$ and on photometric errors. Since the stars plotted in the CMD
of Figure~\ref{f:f8} are the same and the observing conditions of the BHS run
were excellent, the overall appearance of the two CMDs confirms also
the good {\it internal} quality of our photometry. Note that the width 
of the red HB (in particular) is due to the inclusion of randomly phased 
measures of the variable stars.

\smallskip\noindent
$\bullet$ {\it Comparison with Guhathakurta \etal 1995}
\par\noindent
A similar check has been carried out with the GYBS data. Figure~\ref{f:f9}
presents the CMDs for the stars independently detected and measured 
in the common area covered by the two studies (see Figure~\ref{f:f2}, $r 
\simeq $ 1'). In this region, we have found 340 stars while GYBS list 
339 objects. Of them, 15 and 16 stars in the GYBS and in our subsamples,
respectively,  do not have cross-identification. As for BHS, they mostly 
lie near the faint limit of the selected subsets, and the difference
is thus due to the selection bias induced by the photometric scatter.

By inspecting Figure~\ref{f:f9}, it is also interesting to note the
difference in photometric quality achieved in the two samples.
Since the GYBS data have been obtained using the $HST$ WFPC1,
the larger internal scatter is most likely due to the difficulty 
of measuring ``aberrated'' stars in the inner crowded region. On the
other hand, the excellent agreement in the total numbers of selected
objects  confirms the very high degree of completeness achieved in our 
ground-based survey (note that the searching phase is almost ``perfect''
for the bright objects also with $HST$ WFPC1).

\begin{figure}[htbp]
\caption{ Comparison of the upper part of the CMD by GYBS (panel a) with 
our data (panel b) in the spatial region in common to both studies. The
dashed line reports the Horizontal Branch and bright Red Giant and Asymptotic 
Branch limit as adopted in Figure 7 of GYBS.}
\label{f:f9}
\end{figure}

The comparisons discussed above can be considered as {\it external}
independent checks of the degree of completeness actually achieved 
in our searching phase, at least as far as the upper part of the CMD 
is concerned. The results of the comparison have shown that completeness
is close to $100\%$ even using the  $HST$ sample as reference.
Since this comparison refers to the innermost regions 
of the cluster, where completeness is most problematic,
we are confident that a similar very high degree of efficiency has
been guaranteed by our searching routine in the more external fields.
Additional tests made adopting the usual ``artificial star'' procedure fully 
confirm this conclusion.

\smallskip\noindent
$\bullet$ {\it Comparison with Montgomery, 1995:  the bright sample}
\par\noindent
The availability of the new catalog compiled by Montgomery (1995) has offered
the opportunity to carry out a preliminary check of completeness allover our
observed field as his observations fully overlap our considered regions. Since
the limiting magnitudes of the two surveys vary with the cluster regions, we
have separated the analysis of the bright sample from that of the faint one. 

\begin{figure}[htbp]
\caption{ Comparison of the upper part of the CMD by M95 (panel a) with 
our data (panel b) in the spatial region in common to both studies. 
Only stars with $B<18.6$ have been plotted. Variable stars are here 
included.}

\label{f:f9a}
\end{figure}

Figure~\ref{f:f9a} shows the CMDs obtained from the two photometries
down to B$=18.6$, considering just the overlapping area.
As can be seen, apart from a slightly larger scatter in M95 data,
the overall morphology is very similar and also the total number of
stars is comparable. This evidence by itself confirms that the
degree of completeness achieved in the two studies is also similar
and high. We have detected 2841 stars while M95 lists
2750 objects. However, to have quantitative hints of the completeness
of our sample it is useful to check the number and properties of the
objects which are present only in the M95 sample, as they could have
been missed in our search. 

\begin{figure}[htbp]
\caption{ $V$,$B-V$ CMD (panel a) and map (panel b) of the stars detected by
M95 on our area and not cross-identified with objects detected by ourselves.
Coordinates are in arcsec with respect to the cluster center (see Section 
2.3).}
\label{f:f9bis}
\end{figure}

Figure~\ref{f:f9bis} shows the CMD ({\it panel a}) and the map ({\it panel b})
of the objects M95 detected on our field and we apparently missed. It is quite
evident that they are mostly located in the very central regions
where the resolving capability of the two surveys is different,
our own being much better. A careful interactive analysis of
a subset of these objects has shown that most of them are actually
blended images which we resolved into two (or more) separated stars
with off-center positions with respect to the blend. If we increase
the size of the box (from 0.5 to 1.0 arcseconds) we get
multiple identifications with the individual components.

The few bright stars located at the outskirts of the considered field
are lacking in our samples as they are "field" stars according to
Cudworth (1979) or because they are highly saturated standards.

To summarize the comparisons and to give a further insight on the mutual
completeness of the four considered samples, we have plotted in 
Figure~\ref{f:f10new} 
the ratios between the number of stars measured by BHS ({\it panel a}),
GYBS ({\it panel b}), and M95 ({\it panel c}), respectively, and 
the number of stars we measured per bin of 0.5 mag in $V$. The comparison is
obviously made on the areas in common between each quoted study and our own
survey. 
Error bars have been computed 
by simply taking the inverse of the square of the total number of stars 
in each bin. As can be seen the bright samples are substantially the same
especially considering the effect of binning over CMDs
which have slightly different calibrations and local morphologies.

In conclusion, based on the tests here carried out and on those already
reported for the outer samples considered in PH94, we can say 
that most likely all the objects brighter than $V=16.75$ have now been 
detected and measured in a circular annulus with 0.3' $< r <$ 7' in M3.

\begin{figure}[htbp]
\caption{ Comparison of the number of stars we included in the bright samples
with the corresponding lists by BHS (panel a), GYBS (panel b) and M95 (panel c).
The ratios of the numbers of stars we found to those measured by BHS, GYBS, and
M95, respectively, are presented over 0.5 mag bins in $V$.}

\label{f:f10new}
\end{figure}

\smallskip\noindent
$\bullet$ {\it Preliminary comparison with our own HST-data: the bright sample}
\par\noindent
The considerations made above
are sufficient to assess the degree of completeness necessary 
in determining the populations of the bright RGB and AGB (down to
$\sim 1$ mag below the HB). However,  
further discussion is required concerning the completeness of
the faint, blue HB tail which plays a r\^ole in the computation of the
various population ratios we are going to discuss later.

Since we have specifically taken deep UV exposures with $HST$ to
study this aspect and a detailed paper is in preparation,
we simply anticipate here the essence of the results. 
The preliminary reductions of the $HST$ data confirm that their
is no significant population in the HB blue 
tail fainter than $V\sim17$ (for $r > 20"$) except for 
the objects we have detected
with the present BVI CCD-survey and the old PH94 study.

In particular, in the area in common between the present sample and the 
$HST$ survey, we have detected 8 objects in the CCD 
frames lying at $B-V<0.1$, $17.0 < V < 18.5$ and 8 objects in the $HST$ 
frames. Although the detected 
population is small and the membership to the HB blue tail is somewhat 
uncertain given the photometric scatter, this ensures that the loss of 
candidate faint HB stars should be small and in any case quite negligible
in the computation of the global population ratios (see Sect. 3.5).

Outside $r>110"$, where HST-data are not availabe at present, we rest
on the tests for completeness carried out on the photographic plates
used in PH94.  On the other hand, with decreasing significantly the
degree of crowding, the degree of completeness raises quite steeply.

In conclusion, from the available data there is no evidence that the stellar
counts of the HB (down to $B=18.6$, with $B-V<0.1$) 
and the AGB and RGB (down to $V=16.75$) are affected at any significant
level from incompleteness. 
Hereinafter, we will refer to this subset as the {\bf Bright Complete 
Sample (BCS)} to be used for computing useful population ratios
(see Sect. 3.5).

\smallskip\noindent
$\bullet$ {\it Comparison with Montgomery, 1995:  the global sample
over the annulus with $3.5<r<6.0$ arcmin}
\par\noindent
In PH94 we presented a CMD including all the stars we measured within
the annulus with $3.5<r<6.0$ arcmin. The degree of completeness of
that sample was checked using various techniques, but it was impossible
at that time to make any comparison with similar observations in the
same field. We have now the opportunity to directly compare our
data with the M95 sample over the same annulus and get some useful
information.

\begin{figure}[htbp]
\caption{ Comparison of the upper part of the CMD using data from M95 (panel a)
and from the present work (panel b). This comparison is limited to the annulus
between $r=3.5$ and $r=6$ arcmin.} 
\label{f:f9b}
\end{figure}

Figure~\ref{f:f9b} presents the CMds for the objects measured in the
two surveys in the considered annulus. As already noticed for the
bright samples, they are very similar and consistent. Of course,
the total numbers of measured objects are different mostly because
the limiting magnitude we reached is deeper. 

To make a quantitative comparison it is therefore useful to compute
the population ratios over magnitude bins large enough to avoid
small number fluctuations. Figure~\ref{f:f10a} shows the
distribution of the observed ratios as a function of the V magnitudes.
As can be seen, down to V$\sim$18.5 the two samples are essentially
identical and, presumably, ``truly'' complete. For fainter magnitudes,
our own sample clearly  outnumbers the M95 list.

\begin{figure}[htbp]
\caption{ Ratios of the stars detected by M95 and in the present work over the
completeness annulus ($3.5 < r < 6$ arcmin) defined above. Data are binned in
0.5 mag bins in $V$.} 
\label{f:f10a}
\end{figure}

In conclusion, this comparison adds further confirmation that the samples we
secured on the various regions of M3 we observed are sufficiently complete to
be used for testing the ecvolutionary models.

\section{Results }

Figure~\ref{f:radbv} and Figure~\ref{f:radvi} present our CMDs  in 
different radial annuli, and clearly show how the different (CCD and 
photographic) sub-samples have been joined, as well as the main features of 
the individual branches. 
Since the lower part of the $V$, $B-V$ colour-magnitude diagram 
(composed only by re-calibrated photographic data) was discussed at length in 
PH94, we will concentrate here exclusively on the bright part of these
CMDs.

\begin{figure}[htbp]
\caption{ $V$, $B-V$ CMD at different radial distances.}
\label{f:radbv}
\end{figure}

\begin{figure}[htbp]
\caption{ $V$, $V-I$ CMD at different radial distances.}
\label{f:radvi}
\end{figure}
\noindent
The main aspects worth of note are the following:
\begin{enumerate}

\item 
The main branches can easily be delineated in any radial bin,
including the most internal region. In particular, the RGB and the 
AGB can be separated quite easily at the AGB base, located at 
$V \sim 14.9$. The Giant Branch can easily be traced up to the tip, 
which is located at $V=12.63$, $B-V=1.58$ (star $\#$ 4191).
The RGB-bump is
also clearly detectable as a clump of stars at $V \sim 15.45$ (see below). 

\item 
Despite the increased scatter in the CCD sample  due 
to crowding in the inner regions (and to the bright plate limit in the 
B CFHT-exposures just above $B=18.6$), the matching of the different samples 
is adequately smooth. The bulk of stars located just above the 
TO-region within the colour range $0.<B-V<0.65$ are most probably
the result of {\it optical blending} of two bright MS-stars of
similar colour yielding a blend approximately 0.75 mag brighter than
the individual components. This conjecture is further
supported by their progressive disappearance with distance 
from the crowded central regions. The difficulty of properly separating 
these objects represents a crucial problem in the study of the
Luminosity Function of the SGB. As already discussed for instance by
Ferraro \etal (1992a,b), some of these stars may also be
blends of SGB objects and blue stragglers.

\item 
The HB is narrow and, over the considered region 
($0.3'<r<7'$), it contains 186 variable stars, which have not been 
plotted in Figure~\ref{f:radbv} and Figure~\ref{f:radvi}. 
The blue HB tail extends down to $V\sim18.6$,
about half a magnitude brighter than the TO-level, and there is little doubt 
that this extremely blue HB population is quite clearly detached from 
the bulk of the other HB stars by a discontinuity in the population
at $V \sim$16.8--16.9 (see also Section 3.3 in PH94). 
There are also a few stars about half a magnitude brighter than 
the average luminosity of the HB, $<V_{\rm HB}>$, and located 
within 2' from the cluster center. 
While the possibility of field interlopers seems quite improbable
given the high galactic latitude of the cluster, their presence 
could be due to ``optical blending'' between HB and RGB components
or to the existence of a more evolved ($supra-$HB) population.

\item 
Several candidate blue stragglers have been detected and
discussed in F93. Their distribution is better seen in the $V$, $V-I$ CMDs, 
where they trace, as expected, a continuation of the Main Sequence.

\end{enumerate}

\subsection{ Mean ridge lines for M 3}

Normal points for the main branches in the CMD of M 3 (MS, SGB, RGB, HB
and AGB) are presented in Table~\ref{t:ridge}. As usual, mean ridge lines for
each evolutionary phase have been derived by plotting magnitude and colour 
histograms along each branch and by rejecting the most deviating objects 
via a  $k$$\sigma-clipping$. 

\begin{table}
\caption{ Mean ridge lines}
\begin{tabular}{lcccrcc}
\hline\hline
\\
RGB+SGB+MS&    & RGB     &     HB &         &   AGB   &         \\
   V   &   B-V &  V-I    &   V    &    B-V  &     V   &   B-V   \\
\hline
\\
 12.50 & 1.648 &  1.706  &  15.11 &  0.684  &   15.06 &   0.694 \\
 12.70 & 1.509 &  1.554  &  15.18 &  0.671  &   14.87 &   0.733 \\
 12.90 & 1.371 &  1.441  &  15.25 &  0.660  &   14.70 &   0.771 \\
 13.10 & 1.280 &  1.370  &  15.40 &  0.617  &   14.52 &   0.801 \\
 13.30 & 1.208 &  1.311  &  15.56 &  0.573  &   14.38 &   0.831 \\
 13.50 & 1.131 &  1.249  &  15.61 &  0.534  &   14.23 &   0.863 \\
 13.70 & 1.067 &  1.206  &  15.67 &  0.450  &   14.08 &   0.904 \\
 13.90 & 1.019 &  1.169  &  15.67 &  0.400  &   13.93 &   0.939 \\
 14.10 & 0.982 &  1.135  &  15.67 &  0.350  &   13.78 &   0.978 \\
 14.30 & 0.949 &  1.106  &  15.67 &  0.300  &   13.62 &   1.027 \\
 14.50 & 0.919 &  1.076  &  15.67 &  0.250  &   13.46 &   1.084 \\
 14.70 & 0.890 &  1.053  &  15.68 &  0.200  &   13.28 &   1.154 \\
 14.90 & 0.863 &  1.032  &  15.68 &  0.178  &   13.08 &   1.239 \\
 15.10 & 0.840 &  1.009  &  15.70 &  0.158  &         &         \\
 15.30 & 0.820 &  0.984  &  15.72 &  0.144  &         &         \\
 15.50 & 0.803 &  0.963  &  15.75 &  0.129  &         &         \\
 15.70 & 0.789 &  0.948  &  15.80 &  0.110  &         &         \\
 15.90 & 0.770 &  0.932  &  15.86 &  0.092  &         &         \\
 16.10 & 0.768 &  0.913  &  15.92 &  0.073  &         &         \\
 16.30 & 0.757 &  0.899  &  15.96 &  0.064  &         &         \\
 16.50 & 0.744 &  0.886  &  16.00 &  0.057  &         &         \\
 16.70 & 0.733 &  0.869  &  16.26 &  0.020  &         &         \\
 16.90 & 0.722 &  0.858  &  16.53 & -0.023  &         &         \\
 17.10 & 0.712 &  0.850  &  16.79 & -0.060  &         &         \\
 17.30 & 0.702 &  0.838  &  16.94 & -0.072  &         &         \\
 17.50 & 0.692 &         &  17.14 & -0.093  &         &         \\
 17.70 & 0.682 &         &  17.50 & -0.123  &         &         \\
 17.90 & 0.669 &         &  17.88 & -0.153  &         &         \\
 18.10 & 0.648 &         &  18.23 & -0.183  &         &         \\
 18.30 & 0.627 &         &        &         &         &         \\
 18.50 & 0.534 &         &        &         &         &         \\
 18.70 & 0.477 &         &        &         &         &         \\
 18.90 & 0.450 &         &        &         &         &         \\
 19.10 & 0.443 &         &        &         &         &         \\
 19.30 & 0.448 &         &        &         &         &         \\
 19.50 & 0.460 &         &        &         &         &         \\
 19.70 & 0.471 &         &        &         &         &         \\
 19.90 & 0.484 &         &        &         &         &         \\
 20.10 & 0.503 &         &        &         &         &         \\
 20.30 & 0.525 &         &        &         &         &         \\
 20.50 & 0.549 &         &        &         &         &         \\
 20.70 & 0.573 &         &        &         &         &         \\
 20.90 & 0.598 &         &        &         &         &         \\
 21.10 & 0.629 &         &        &         &         &         \\
 21.30 & 0.664 &         &        &         &         &         \\
 21.50 & 0.700 &         &        &         &         &         \\
 21.70 & 0.737 &         &        &         &         &         \\
 21.90 & 0.773 &         &        &         &         &         \\
\\
\hline
\end{tabular}
\\
\label{t:ridge}
\end{table}

\noindent
Specifically concerning the determination of \vhb, we have computed a running
mean over a 0.2 mag box moving along the HB in colour; 
this procedure yields a value 
\vhbm$=15.66\pm0.03$ which represents the average magnitude level
of the HB, thus including the effects due to evolution off the Zero Age
HB (ZAHB) within the instability strip. 

This value is the mean level of the HB obtained from the constant stars (at the
edges of the instability strip). The mean magnitudes of the RR Lyrae variables
within the strip will be discussed elsewhere. The associated uncertainty is the
observed {\it rms} vertical scatter of the HB at that colour. 

Although evolution off the ZAHB is short compared to the lifetime of 
the ZAHB-phase (i.e. that spent at almost constant luminosity,
see f.i. Sweigart and Gross 1976, Sweigart \etal 1987, Lee \etal 1990),
one should quite carefully distinguish between the {\it average}
HB luminosity and the ZAHB luminosity. There are essentially two
ways to take this difference into account: the first,
by adding a small positive correction to \vhbm ($\sim0.03-0.05$ mag) to
compensate for evolution; the second, by adopting as \vzahb\ 
the {\it lower} envelope of the HB distribution in luminosity
within the instability strip (corresponding at Log T$_e\sim3.85$).
Using our HB sample, we would get \vzahb= 15.70 in both 
approaches.

Since all these effects are actually partially smeared out by the photometric
errors, the difference between the ZAHB and the average levels
is only marginally significant. In conclusion, we
will adopt \vhbm$=15.66\pm0.03$ and \vzahb$=15.70\pm0.03$.

\subsection{ The metallicity of M 3: a major change?}

Based on the results presented in the previous Sections, we can now 
derive a new estimate of the metal abundance of M 3 using the so-called 
{\it photometric} indicators.

The main photometric parameters related to the cluster mean metallicity
are \bmvg\ (Sandage and Smith 1966) and \d14\ (Sandage and Wallerstein 1960).
Adopting \vzahb=15.70, $E(B-V)=0.00$ (see Table 10 of PH94),
and the mean ridge line of the RGB listed
in Table~\ref{t:ridge}, we obtain \bmvg $=0.80\pm0.03$ and \d14 $=2.81\pm0.03$.
Using then the relationships  quoted in Table~\ref{t:meta} we derive
values for [Fe/H] ranging from $-1.68$ to $-1.45$~dex.

\begin{table}
\caption{ Metallicity of M 3 from the RGB photometric parameters}
\begin{tabular}{llc}
\hline\hline
\\
Calibration                  & References              & [Fe/H] \\
	                     &                         &        \\
\hline
\\
4.30$[B-V]_{o,g}$-5.00       & Zinn $\&$ West 1984     & $-1.56$ \\
3.84$[B-V]_{o,g}$-4.63       & Gratton 1987            & $-1.56$ \\
4.68$[B-V]_{o,g}$-5.19       & Costar $\&$ Smith 1988  & $-1.45$ \\
2.85$[B-V]_{o,g}$-3.76       & Gratton \& Ortolani 1989& $-1.48$ \\
\\
\hline
\hline
\\
-0.924$\Delta V_{1.4}$+0.913 &Zinn $\&$ West 1984      & $-1.68$ \\
-1.01$\Delta V_{1.4}$+1.30   &Costar $\&$ Smith 1988   & $-1.54$ \\
-0.65$\Delta V_{1.4}$+0.28   &Gratton \& Ortolani 1989 & $-1.54$ \\
\\
\hline
\end{tabular}
\\
\label{t:meta}
\end{table}
\noindent
Another {\it photometric} estimate of the metal abundance of M~3 can
be obtained from the CMD in the ($V,~V-I$)-plane by exploiting the iterative 
method recently defined by Sarajedini (1994). 
From the mean ridge line listed in Table~\ref{t:ridge}, we obtain 
$(V-I)_{o,g}=0.951$, $\Delta V_{1.2}=$ 1.82 $\pm$ 0.14, $E(V-I)=$ 
0.00 $\pm$ 0.05,  and, eventually, [Fe/H]$=-1.45$ with a formal error 
of $\pm0.17$~dex. 

From Table~\ref{t:meta}, the lowest metallicity value is the same as 
the widely used figure
from Zinn and West (1984; [Fe/H]$=-1.66$), and is based on their
metallicity-scale calibration obtained via integrated cluster
observations.

The higher value ($-1.45$) is based on direct spectroscopic 
determinations of [Fe/H] from individual stars,
as opposed to the integrated indexes used by Zinn and collaborators.
The latest high resolution spectroscopic investigations, namely 
Sneden {\it et al.} (1992; SKPL) and Carretta 
and Gratton  (1996a; CG96), show that the problem of the 
reliable measure of the metal abundance of this template cluster is 
not trivial at all.

In fact, SKPL found \fe$=-1.46$ from the analysis of 7 stars
(and a slight different value of -1.42 by adding 3 other stars, see Kraft 
\etal 1995), and CG96 derived \fe $=-1.34 \pm 0.02$ ($\sigma=0.06$) from
10 stars.
Both studies are based on high resolution, high signal-to-noise CCD echelle
spectra of high quality. In particular CG96 have re-analysed the
equivalent widths published by SKPL for M 3 and other clusters to obtain 
a new, homogeneous metallicity scale for 24 calibrating clusters. 

The main differences between the two quoted studies (apart from minor 
changes in the adopted values for the microturbulent velocity) are in the 
set of adopted atomic parameters (in particular oscillator strengths $gf$ 
for Fe~I and Fe~II) and in the choice of the model atmospheres used in 
the analysis. CG96 used the latest, updated models from the grid of Kurucz 
(1992), that allow an homogeneous comparison between solar and stellar 
abundances, alleviating a major drawback 
of any former analysis of abundances for 
globular cluster stars. 

In fact, as discussed by CG96, all previous spectroscopic determinations
of the [Fe/H] content of globular cluster stars were actually systematically
uncertain because of the large differences ($\sim 0.15$~dex) still
existing in the reference solar value adopted in the absolute analyses. 
In particular, a significant discrepancy persists in the solar Fe abundances 
as obtained using the Holweger and M\"uller (1974--HM) semi-empirical solar 
model (usually considered to be the best reproduction of the solar atmosphere)
and the model atmospheres proposed by Bell \etal (1976), generally
adopted (also in the SKPL papers) in the analysis of cool cluster giants (see
also Leep et al. 1987). The crucial problem thus is to find a firm
answer to the question: {\it what is the correct solar abundance to adopt
as reference for Fe?}. 

The recent analysis by CG96 has apparently settled the discrepancy as 
they have obtained a revised determination of the solar Fe abundance
which is very similar to the value given by the HM model. This result
has been obtained by  using the
set of $gf$ values discussed in CG96 and a solar model extracted 
from the same grid of Kurucz (1992) models as used for the analysis of 
the cluster giants. 
Consequently, the study of CG96 yields a systematic difference, $\Delta$\fe 
$=+0.12\pm 0.01$ ($\sigma$=0.08 over 162 stars analyzed; 0.08 in the case 
of M 3) with respect to previous analyses, which used the Bell \etal 
(1976) models that are $\sim 150$~K cooler than HM in the line formation 
region.
Therefore, even if the observational material ($i.e.$ the equivalent 
widths) is the same, the analysis of CG96 seems to be more self-consistent 
than that of SKPL.

Since the precise determination of the metallicity of M 3 has important
implications on various items (f.i. on the long-standing problem of
the so-called Sandage Period Shift Effect, see Sandage 1993),
it may be useful to analyse further the discrepancy between the
value \fe$=-1.66$ obtained via the Zinn and West (1984) calibration
and the significantly higher \fe$\sim -1.4$ obtained from 
high-resolution spectroscopy. 

Zinn and West (1984) based their estimates on integrated cluster features
ultimately calibrated using old (and sometimes uncertain) \fe abundances
obtained from photographic high-dispersion spectra (Cohen 1983). 
However, from more than 160 giants homogeneously analyzed 
in 24 calibrating clusters, CG96 have demonstrated that the 
ZW metallicities differ significantly from these new results. 
In particular, they are about 0.10~dex higher for [Fe/H]$>-1$, 0.23~dex 
lower for $-1<$[Fe/H]$<-1.9$, and 0.11~dex too high for [Fe/H]$<-1.9$.
This non-linearity of the ZW scale is significant at 3$\sigma$ level and 
cannot be ignored when discussing astrophysical problems involving tiny
metallicity  differences among the clusters.

On the basis of the CG96 new scale, Carretta and Gratton (1996b) have 
also derived a new calibration of \bmvg\ in terms of \fe. Using the values 
of \bmvg\ from the compilation of Michel and Smith (1984) for
22 calibrating cluster on the new scale, they obtained a $2^{nd}-$order 
polynomial fit that closely resembles the theoretical calibration of 
Demarque \etal (1982) from evolutionary red-giant models. 
Based on this new calibration, the metallicity of M 3 turns out to be
\fe$=-1.38$, with a 1$\sigma$ error bar of 0.18~dex, 
in good agreement with the spectroscopic determination 
reported by CG96.

In conclusion, from the detailed re-analysis discussed above, we are
inclined to believe that the metallicity of M 3 is slightly
larger than estimated so far and probably the best value to adopt
at present is \fe$=-1.45\pm0.10$, where both the absolute figure and the
size of the error are just the result of our global overview.

\subsection{ The bump on the Red Giant Branch }

Theoretical evolutionary models predict the existence of a special
feature along the RGB called the ``RGB-bump'' (see e.g. Thomas 1967, Iben 
1968, RFP88). As discussed in detail among others by
Rood and Crocker (1985), the practical detection of such a feature 
requires the availability of very populous RGB samples. 
Fusi Pecci \etal (1990) detected the RGB-bump in 11 globular
clusters, including M 3 (based on our early PH94 data; see also Ferraro 1992). 
The use of our new data-base allows us to confirm the detection of the
RGB-bump (see Figure~\ref{f:f4}, Figure~\ref{f:bumptot})
located at $V_{bump}$=15.45$\pm$0.05, i.e. exactly  the same value as
the previous measurement 
listed in Table 4 of Fusi Pecci \etal (1990), and we thus refer
to that paper for the discussion of this specific issue.

\begin{figure}[htbp]
\caption{ Integrated (panel a) and differential (panel b) luminosity functions
of the RGB (Bright Complete Sample). The vertical line in panel (b) shows the
location of the $RGB-bump$.} 
\label{f:bumptot}
\end{figure}

\subsection{ The HB population and morphology in M 3 }

Based on our bright sample, containing $all$ the stars brighter 
than $B=18.6$ and located
within a square of about 7'$\times$ 7' (but with $r>20"$), 
the HB of M 3 spans a very wide colour range ($\sim 1$ mag), from the red 
end at $B-V\sim$0.7 to the bluest stars at $B-V\sim-0.3$. Therefore 
the distribution in effective temperature is very wide, and, in turn, 
there is a wide 
spread in the HB mass distribution (Rood and Crocker 1985). 

Recently, Fusi Pecci \etal (1992, 1993) have discussed in detail the 
properties of the observed HB morphologies (and in particular the so-called
"Second Parameter" problem) in relation to stellar mass loss, the effect 
of the environment on the evolution of the individual stars,  and 
the presence of binary systems (primordial, collisional, merging, etc.). 
In particular, the HB of M 3 has been 
{\it dissected} into sub-groups having presumably
different evolutionary histories. The basic idea is that both the
blue and the red extremes of the observed HB might include 
peculiar objects which are intrinsically different
from {\it genuine} HB stars and are rather the result of the 
stellar and dynamical evolution of binary systems (at least in part
related to the blue stragglers), or are HB objects which keep track
of interactions causing an ``extra-mass loss'' from the envelope
during the previous evolutionary stages.

For the sake of brevity we refer to the quoted papers for a complete
discussion and simply re-analyse the content of the boxes which have 
been defined in PH94 to describe the HB morphology. 

\begin{figure}[htbp]
\caption{ Enlargement of the colour-magnitude diagram of M 3 in the HB region,
with boxes defining the areas we identified along the HB. All stars belonging
to our BCS sample are plotted in this Figure apart from the 186 known
variables (see text). }
\label{f:f11}
\end{figure}

\noindent
As show in Figure~\ref{f:f11} we have ``dissected'' the HB into seven 
boxes and report below a few notes on each sub-group.
\begin{enumerate}

\item 
{\it Group HEB}: 10 stars (2 in PH94), 
the faintest one reaching $V \sim 18.3$ and with $B-V<-0.1$, but 
a few other bluer and/or fainter objects might deserve a further check.
Their location in the CMD could well be due to large  photometric errors
or they could be non-members. Most of them are confined 
within the inner 2 arcmin from the cluster center. If their photometry
and membership are confirmed, they could represent good candidates
to search for the possible descendants of late collisional mergers 
(Bailyn and Iben 1989, Fusi Pecci \etal 1992) or they could be post-HB 
stars (such as AGB-manqu\`e, see e.g. Greggio and Renzini 1990, Cacciari 
\etal 1995, D'Cruz \etal 1996). 

\item 
{\it Group EB}: 10 stars (4 in PH94). Together with the 10 HEB 
stars, this group constitutes a subset of 20 stars clearly segregated from 
the bulk of the HB stellar population. Their total number and location seems 
sufficient to exclude the possibility  that all of them are interlopers. 
Hence, we are inclined to conclude that there is now clear-cut observational 
evidence for the existence of a very extended (though sparsely populated)
blue HB tail in M 3, possibly separated by a gap from the bulk of the 
{\it normal} HB objects. The $HST$ UV data currently being analyzed, 
coupled with spectroscopy, could be very helpful
in understanding their true origin.

\item 
{\it Group B}: 206 objects, (70 in PH94). The population of 
this box includes presumably {\it normal} HB stars located above a small 
(apparent?) gap at $V\sim 16.9$ and bluer than the blue edge of the 
instability strip. The scatter of the points around the mean ridge
line is compatible with the size of the observational
errors (note that the B exposures are short), but it could also 
partially reflect the effects due to evolution off the ZAHB for some
stars.

\item 
{\it Group V}: 186 stars (85 in PH94). This group contains 
only RR  Lyrae variables.
In Figure~\ref{f:f11} variables are not reported as appropriate
mean values are still lacking. At present, mean $V$ and $B-V$ values 
are listed for the 85 HB variables of the photographic sample
in Table 10 of PH94. For the other 100 HB variables 
mean magnitudes and colours are not available from the present photometric 
data. Consequently, their presence in the plotted CMD would introduce
a ``confusing'' scatter of the points both in magnitude and in colour 
mostly due to a random-phase effect of the available
measurements, which are insufficient to cover completely the light curve. 

A few of these stars  could however also be evolved HB stars brighter than the
ZAHB. Finally, there are 17 constant stars which are located within the box
because of photometric errors or due to the existence of some overlapping in
colour of the variable and non-variable strips. This specific item will be
dealt with in more details in a forthcoming paper. However, we have included
them in the total number of HB stars. 

\item {\it Group R}: 106 stars (51 in PH94). They lie to the red of the 
instability strip and reach up to about half a magnitude brighther and
redder than the ZAHB level within the instability strip.

\item 
{\it Group ER}: 15 stars (7 in PH94). This group overlaps in 
colour with the previous one, but it is composed of stars brighter than 
$V\sim 15.3$ and still well separated from the AGB base. As mentioned in
PH94, at least some of them could be the ``progeny'' of BSS (see Fusi Pecci 
\etal 1992) and they are thus worth of further study.

\item {\it Group SHB}: 7 stars. They are located
well above the HB and could represent {\it highly evolved} HB objects
which are travelling from their original ZAHB location towards the
AGB. Alternatively, they could be blends  of HB+RGB stars.

\end{enumerate}

\subsection{ Star counts and population ratios: mixing and He abundance }

A complete sample of stars with accurate photometry offers a major tool for
an observational verification of the predictions of the stellar evolution 
theory. In particular, star counts in a given evolutionary phase yield
a direct test of the duration of that specific phase (Renzini and Buzzoni 1986).
In turn, this allows one to get a deeper insight on unsolved problems like 
for instance the extent of the mixing phenomena, or to measure important
basic quantities like the primordial helium abundance Y (see RFP88).

The approach to the problem is well established (see e.g. RFP88, 
PH94): from the CMD one can measure a set of parameters to be used along with 
the appropriate calibrations based on theoretical models. 
The most frequently used are:
\begin{itemize}
\item $R=N_{HB}/N_{RGB}$, the ratio of the numbers of stars on the HB 
and on the RGB brighter than ZAHB luminosity at Log$T_e\sim3.85$;
\item $R'=N_{HB}/(N_{RGB}+N_{AGB})$, which includes also  the number 
of stars in the AGB phase;
\item $R1=N_{AGB}/N_{RGB}$;
\item $R2=N_{AGB}/N_{HB}$.
\end{itemize}
\noindent
To preserve full homogeneity with the study of Buzzoni \etal
(1983, BFBC) which is still the standard reference for this subject, 
we have adopted their prescriptions. In particular, we fixed 
the separation between HB and AGB at $V\simeq 14.9$, about 0.8 mag above 
the mean magnitude of the HB, and counted RGB stars brighter than 
$V=15.81$, adopting a differential bolometric correction 
$\Delta$B.C.=0.11 mag (O. Straniero, private communication) between HB and RGB
stars.
The results for the star counts along the various branches and the 
derived parameters are listed in Table~\ref{t:ratios}.

\begin{table}
\caption{ Star counts on HB, AGB, and RGB of M 3}
\begin{tabular}{ll}
\hline\hline
\\
N$_{RGB}$ & $=$ 414 \\
N$_{HB}$  & $=$ 550 \\
N$_{AGB}$ & $=$ 69  \\
$R$       & $=$ 1.33 $\pm$ 0.12 \\
$R'$      & $=$ 1.14 $\pm$ 0.10 \\
$R1$      & $=$ 0.17 $\pm$ 0.03 \\
$R2$      & $=$ 0.12 $\pm$ 0.02 \\
\\
\hline
\end{tabular}
\\
Note: all counts and ratios refer to our Bright Complete Sample (BCS).
\label{t:ratios}
\end{table}

\noindent
The parameters $R1$ and $R2$ essentially
indicate the relative ratios of the lifetimes of the AGB, RGB, and HB.
Since these lifetimes are basically driven by nuclear burning and
mixing phenomena (which may alter the ``local'' fuel quantity), different
ratios are predicted from the models with varying mechanisms and
extension of mixing for fixed properties of the nuclear burning.
Within the framework discussed by RFP88,
the values here obtained for $R1$ and $R2$ are fully compatible
with the predictions of ``standard and canonical'' models.

The values for $R$ and $R'$ can be used along with eqs. 11 and 13 of 
BFBC to obtain an estimate of the primordial He abundance based on their 
calibrations of these parameters in terms of Y.
From the values listed in Table~\ref{t:ratios} (the errors are just from 
count statistics), we obtain Y$_R = 0.22 \pm 0.02$ and Y$_R' = 0.23 \pm 0.02$, 
in excellent agreement with the mean value found by BFBC as well as with 
other estimates (see e.g. Boesgaard and Steigman 1985, 
Olive and Steigman 1995 and ref. therein).
In the computation we have considered for the HB
all the stars included in the samples ER+R+V+R+EB+HEB. This may not be
fully correct if at least some of them are {\it non-genuine} HB
stars (in the sense that they are not evolving "directly" from the RGB).
However, since their number (including the few SHB stars) is so small,
the result is practically unaffected taking into account the size of
the errors. For instance, by deleting from the HB all the
stars in the samples ER+EB+HEB, the computed ratios would
become $R=1.24$, $R'=1.07$, $R1=0.17$, $R2=0.13$, which would leave the 
conclusions on $Y$ unchanged. 

\subsection{ The Blue Stragglers population of M 3: toward the definition of
a complete and reliable sample}

A comprehensive discussion on the BSS population of M 3 has already been
presented in F93. Therefore, here we shall only 
extend to the BSS region the same kind of comparisons with the data-sets
of other previous studies in order to verify the quality of the
different samples in terms of completeness and selection bias.
We anticipate that the problem of identifying a truly pure sample of
BSS candidates is far from being solved, in particular in the inner regions
of the clusters where crowding undermines photometry and for the
faint BSS, whose separation from {\it normal} TO stars is always
difficult and somewhat subjective.

We report below identifications and comparisons made with respect to 
BHS for the region in common with our newly re-calibrated CFHT 
CCD data. The comparison will be based on $V$ and $I$ colours only,
because of the poor response of the chip used for the observations in the
$B$ filter.

Figure~\ref{f:bssp20} shows the result of the comparison with the sample
obtained by BHS for $r>20"$. {\it Panel (a)} shows the data from 
BHS (kindly made available to us by Dr. M. Bolte) with their instrumental 
magnitudes shifted to our system. The box reproduces 
the polygonal region used by BHS (their Figure 3) for selecting their BSS 
candidates. Open squares are the stars in the BHS sample that are the 
counterparts of our BSS candidates on the BHS field. {\it Panel (b)} displays
our CCD-sample, restricted to the BHS field. The box is the same as we used to 
delimitate the region occupied by our (F93) BSS candidates and is identical 
to that in Figure 3 of F93. 
Open squares are the stars of our sample which are counterparts of 
the BSS candidates selected by BHS and falling outside our adopted limits.
In both panels, black squares represent stars that are considered BSS by both
surveys.

As can be seen, there are only 15 BSS candidates in
common (the black squares), while there are 50 stars
labeled as BSS candidates in only one of the two surveys. 
Could all of them be mis-selected; 
could they actually be {\it normal} stars? The answer is difficult to find,
but some additional considerations may be useful.

Our 25 BSS candidates not identified as such by BHS mostly fall
between their BSS box and their SGB. Judging on the basis of
the scatter in the TO region, our photometry seems to display 
smaller internal errors. This may indicate that we can better separate
the two populations because of the different quality of the photometry. 
Another obvious explanation is, of course, that we reached too close 
to the MS in defining our BSS box. 
On the other hand, the 10 stars identified as BSS candidates only by BHS
fall in {\it panel (b)} of Figure~\ref{f:bssp20} at somewhat brighter 
magnitudes than the TO region. Since the quality of the CCD frames
available to BHS seems to be better as far as seeing conditions are
concerned, this might imply that we were unable to separate the
optical blends formed by a BSS-candidate and a SGB star.
$HST$-data should be suitable to settle the issue.

\begin{figure}[htbp]
\caption{ Comparison of the BSS found in our sample and in BHS study. 
In panel (a) are
plotted the data from BHS, with instrumental magnitudes shifted to our system.
The box reproduces the polygonal region used by BHS (their Figure 3) for 
selecting their BSS candidates. Open squares are the stars in BHS sample that
are the counterparts of our BSS candidates on the BHS field. Panel (b) shows
our CCD sample, restricted to the BHS field; the box is the same we used to 
encircle the region occupied by our BSS candidates (Figure 3 of F93). 
Open squares are the stars of our sample which are counterparts of BHS
BSS candidates. 
In both panels, black squares represent stars that are considered BSS by both
surveys.}
\label{f:bssp20}
\end{figure}

\section{Summary and Conclusions}

We have presented the BVI CCD data for about 10,000 stars in the region 
0.3' $< r < \sim$ 4' of M3. These data, along with our previous photographic 
photometry of additional $\le$ 10,000 stars in the external regions as far 
as 7', provide a homogeneous and popolous data base for detailed population 
studies. Completeness has been achieved for all objects with V$\le$18.6. 

Our main results are: 

\begin{itemize} 

\item
A new absolute photometric calibration has been obtained. The comparison 
with the absolute calibration adopted so far for the same field
(Sandage and Katem 1982) shows that there is a significant colour term, in 
the sense that our calibration is $\sim$0.07 mag redder 
than Sandage and Katem (1982) in the blue range e.g. at B-V=-0.2, slightly 
redder ($\sim$ 0.03 mag) at the instability strip, and $\sim$0.08 mag bluer 
in the reddest part of the CMD, e.g. at B-V=1.6. This has important 
implications in the photometric determination of the metallicity, and in 
other items such as the period-shift effect which will be discussed in a 
forthcoming paper. 

\item
After completion of our reductions, we became aware of the availability 
of a new independent photometry of a very wide sample of M3 stars 
(about 23,700) carried out by K.A. Montgomery (1995 -- M95),
fully overlapping our considered regions.
The use of this new catalog has also allowed us to get a link to 
the photometry presented by Stetson and
Harris (1988) for an external area, outside our surveyed region.
In synthesis, there are now three independent calibrations referred 
to observations of Landolt's (1992) standards. They are in marginal 
agreement, but it is difficult at the present stage to choose which 
is the best.
Since the calibration here adopted seems to be compatible with
that obtained by Stetson and Harris (1988) (they actually show
a systematic difference of 0.04 both in B and V, yielding however no colour
difference), we decided to keep using our new calibration.

\item
Based on these new data and calibration, the metallicity derived from 
the photometric indicators $(B-V)_{0,g}$ and $\Delta V_{1.4}$ is 
[Fe/H]$\sim$-1.45. This value is significantly higher than the 
value -1.66 commonly used for M3, and is in very good agreement with the 
most recent metallicity determinations from high resolution spectroscopy 
of individual stars. 

\item
Evidence has been found of the presence of a faint and blue extension 
of the HB, although scarcely populated. These stars, however, could be 
non-genuine HB members but rather the results of merging or collisions. 
More photometric and spectroscopic work is needed in order to assess the 
true nature of these objects. 

\item
From the population ratios of HB, RGB and AGB stars we find a value 
for the helium content Y=0.23$\pm$0.02. This confirms the previous 
determinations of helium content in globular clusters by this method. 

\item
A significant population of blue straggler stars has been detected and 
identified. However the exact number of BSS must await HST data for a 
better space resolution in the most internal regions, for resolving 
optical blends that may turn pairs of TO stars into BSS, and in general for 
a better photometric accuracy. Spectroscopic follow-up studies will be 
extremely important for a better understanding of the formation mechanisms 
of these stars. 

\end{itemize}

\begin{acknowledgements}
We wish to thank H. Richer and G. Fahlman for their collaboration in 
taking the obervations at the CFHT. We are grateful to the CFHT and the 
Calar Alto Observatories for the kind hospitality and support provided during 
the observations. We are specially indebted to K. Janes for providing us 
with his independent sample of M 3 stars in advance of publication, which 
has allowed us to perform important tests and corrections, and M.Bolte for
providing us with his data on Blue Stragglers stars. 
We are pleased to acknowledge R.T. Rood for his critical 
reading of the manuscript and several useful comments and suggestions. 
This work has been supported by the Italian Ministry for Research (MURST). 
\end{acknowledgements}

\end{document}